\documentclass[superscriptaddress,nobibnotes,aps,prd,showpacs,nofootinbib]{revtex4}
\usepackage{amsmath}
\usepackage{amsfonts}
\usepackage{amssymb}
\usepackage{graphicx}

\usepackage{epsfig}

\begin{document}

\title{Soliton models for thick branes}
\author{Marzieh Peyravi}
\email{marziyeh.peyravi@stu-mail.um.ac.ir}
\affiliation{Department of Physics, School of Sciences, Ferdowsi University
of Mashhad, Mashhad 91775-1436, Iran}
\author{Nematollah Riazi}
\email{n_riazi@sbu.ac.ir}
\affiliation{Physics Department, Shahid Beheshti University, Evin, Tehran 19839, Iran}
\author{Francisco S. N. Lobo}
\email{fslobo@fc.ul.pt}
\affiliation{Instituto de Astrof\'{\i}sica e Ci\^{e}ncias do Espa\c{c}o, Faculdade de
Ci\^encias da Universidade de Lisboa, Edif\'{\i}cio C8, Campo Grande,
P-1749-016 Lisbon, Portugal}

\begin{abstract}
In this work, we present new soliton solutions for thick branes in $4+1$ dimensions. In particular, we consider brane models based on the sine-Gordon ($SG$), $\varphi^{4}$ and  $\varphi^{6}$ scalar fields, which have broken $Z_{2}$ symmetry in some cases, and are responsible for supporting and stabilizing the thick branes. The origin of the symmetry breaking in these models resides in the fact that the modified scalar field potential may have non-degenerate vacuua. These vacuua determine the cosmological constant on both sides of the brane. We also study the geodesic equations along the fifth dimension, in order to explore the particle motion in the neighbourhood of the brane. Furthermore, we examine the stability of the thick branes, by determining the sign of the $w^2$ term in the expansion of the potential for the resulting Schrodinger-like equation, where $w$ is the 5-dimensional coordinate. It turns out that the $\varphi^4$ brane is stable, while there are unstable modes for certain ranges of the model parameters in the SG and $\varphi^6$ branes.
\\


\end{abstract}

\date{\today }
\maketitle
\section{Introduction}

Since there is no known fundamental principle requiring spacetime to be $(3 + 1)-$dimensional \cite{Maartens:2003tw,Ge2}, it has been suggested that our observable universe might be a $(3+1)-$dimensional brane in a higher dimensional space \cite{lin,PCU,Ge3}. In most models, there are one or more flat 3-branes embedded discontinuously in the ambient geometry \cite{de}. Moreover, ideas with two 3-branes provide a very elegant description of the large hierarchy between the scales of weak and gravitational forces \cite{de,CG} and contain massless modes which reproduce Newtonian gravity at large distances on the brane \cite{de}. In recent years, particle physics extra-dimensional theories beyond the standard model have become a standard part of the array of phenomenological models \cite{Ge1,Ge5}. Although there are still no experimental evidence supporting extra dimensions, due to various theoretical motivations, extra dimensional models continue to be widely considered in the literature \cite{blg}.
In this context, it would be quite useful to have a set of simple and sufficiently general rules which would allow one to test new models \cite{lin}. Most extra-dimensional models require the existence of scalar fields, for instance, to generate a domain-wall which localizes matter fields \cite{Ge2}. The scalar fields also serve to stabilize the size of the compact extra dimensions \cite{lin,PCU}, and can also help modify the Randall-Sundrum warped-space \cite{Randall:1999ee,Randall:1999vf} to a smoothed-out version \cite{Ge3,de}, or to cut off the extra dimension at a singularity \cite{CG,Ge1}. In order to replace an infinitely thin brane with a thick one, a scalar field with soliton behavior is frequently invoked. The nonlinearity in the scalar field and in particular the existence of discrete vacuua in the self-interaction of the scalar field lead to the appearance of a stable localized solution, which is a good motivation for building thick brane models \cite{blg}. A  general method for determining the lowest energy configuration has been worked out in \cite{Ge4,br}.

Recently, braneworld models have also been considered in higher order curvature gravity and in modified teleparallel gravity. For instance, five-dimensional modified teleparallel gravity was considered in a brane scenario, where analytic domain walls were found to have a double-kink solution in the aftermath of the torsion of spacetime \cite{JY}. Furthermore, this model was extended by using a first-order formalism to find analytical solutions for models that include a scalar field with standard and generalized dynamics. In addition to this, it was found that the brane splits, as a result of the deviation from the standard model by controlling specific parameters \cite{RM}.
Modified gravity in five dimensional spacetime has also been analysed in the Palatini formalism. For instance, a thick Palatini $f(R)$ brane described by an anti-de Sitter warped geometry with a single extra dimension of infinite extent, sourced by a real scalar field was studied in a perturbative scenario \cite{DBL}. Besides, the model of a domain wall (thick brane) in a non-compact AdS space time with only one extra dimension was further analysed in \cite{OON,Andrianov:2012ae}.
The classical tests of General Relativity in thick branes were also studied by studying the motion of test particles in a thick brane scenario and the impact of the brane thickness on the four-dimensional path of massless particles was explored in \cite{FD}. More specifically, by applying a confinement mechanism of massive tests particles in the domain wall, for instance, that simulates classically the trapping of the Dirac field in a domain wall, the influence of the brane thickness on the four-dimensional (4D) path of massless particles was analysed.
A generalized version of the Randall-Sundrum II model with different cosmological constants on each side of a brane were also discussed, where specific configurations of a scalar field and its stability as a replacing factor of the singular brane were considered \cite{ALB}. Models of thick branes in noncompact five-dimensional bulk with different anti-de Sitter geometries on each side of the brane were explored \cite{AA,Andrianov:2013vqa}, and cosmological applications of soliton-like thick branes have also been studied \cite{Ahmed:2013lea}. On the other hand, the existence of brane solutions were also considered as a result of a real scalar field in the presence of five dimensional $f(R)$ gravity \cite{AB}. In addition to this, asymmetric thick branworld scenarios were studied, by changing the superpotential of the scalar field \cite{BM}.

It is widely practiced that in brane world scenarios $Z_{2}$ symmetry is assumed \cite{blg,AIS,DM,CC}, which is originally motivated from the $Z_{2}$ symmetry considered in M-theory \cite{AIS}. Under this symmetry the bulk metric on the two sides of the brane should be the same \cite{CC}. Moreover, under such a symmetry the empty bulk on either sides of the brane have the same negative cosmological constant and as a result they are AdS \cite{DM}. Note that these conditions are satisfied in the Randall-Sandrum model.
There are, however, brane models in which there is no $Z_{2}$ symmetry and the bulk is different on both sides of the brane \cite{AIS}. In the latter, the Friedmann equation for a positive brane tension situated between two bulk spacetimes that posses the same 5D cosmological constant, but which does not possess a $Z_2$ symmetry of the metric itself was derived, and the possible effects of dropping the $Z_2$ symmetry on the expansion of our Universe were examined. In some of these models, the cosmological constant differ on both sides of the brane \cite{JL}, where the effects of including a Gauss-Bonnet combination of higher-order curvature invariants in the bulk action are taken into account. In fact, by considering braneworld scenarios including the Gauss-Bonnet term, it was found that the cosmological dynamics have the same form as those in Randall-Sundrum scenarios but with time-varying four-dimensional gravitational and cosmological constants \cite{CC}.
Motivated by such a possibility, we consider in this paper several models, namely, the sine-Gordon ($SG$), $\varphi^{4}$ and $\varphi^{6}$ brane models which have broken $Z_{2}$ symmetry in some cases. What is meant in the paper as $Z_{2}$ symmetry is the symmetry with respect to the position of the brane (not the actual symmetry in the lagrangian).  The brane position is shifted from $z=0$ for some models. As a result, we find that for the $\phi^6$ and SG systems this symmetry is broken and the vacua on the two sides of the brane are not degenerate. In several cases, the $Z_{2}$ symmetry can be restored by a proper choice of model parameters. The origin of symmetry breaking in our models resides in the fact that the modified scalar field potential may have non-degenerate vacuua. These vacuua determine the cosmological constant on both sides of the brane.

Relative to the stability issue, topological solitons are known for their non-singular structure  and a natural localization mechanism which are highly stable. Zeldovich {\it et al.} \cite{br,81} suggested that the soliton of the $\varphi^4$ model is a reasonable source for the formation of domain walls. Vilenkin extended this idea to incorporate the general theory of relativity \cite{br,86,87,88}. Since there is a close similarity between domain walls in 3+1 dimensions and branes in 4+1 dimensions, it is natural to think that the soliton idea might have something to do with the existence and stability of branes. Motivated by this idea, we consider soliton models for thick branes and extend some of the existing works.
In particular, in \cite{Brihaye:2008am} the authors have studied the stability of the $\varphi^4$ kink brane model in five dimensions, as well as the $\varphi^3$ and the inverted $\varphi^4$ potential. Furthermore, the properties of fermions coupled to the sine-Gordon brane model were investigated. In \cite{Ahmed:2012nh}, double kink-like solutions were considered and the stability of scalar, vector and tensor perturbations were discussed.

This paper is outlined in the following manner: In Section \ref{11}, we briefly review the thick brane formalism, by presenting the action and the field equations. Furthermore, we also study the geodesic equations along the fifth dimension, in order to explore the particle motion in the neighbourhood of the brane. In Section \ref{22}, we present new soliton models and discuss their fundamental properties, for instance, by exploring the broken $Z_2$-symmetry character of the solutions and the confining effects of the scalar field on the brane. In Section \ref{33}, we analyse the stability of these brane models, where the metric and the scalar field are perturbed about the static brane and the resulting equations of motion are linearized in the proper gauge. Finally, in Section \ref{Concl}, we present our concluding remarks.

\section{Thick Brane Formalism}\label{11}

We consider a thick brane, embedded in a five-dimensional (5D) bulk spacetime, modelled by the following action
\begin{equation}\label{ac}
S=\int
d^{5}x\sqrt{|g^{(5)}|}\left[\frac{1}{4} R[g^{(5)}]-\frac{1}{2}\partial_{A}\varphi\partial^{A}\varphi-V(\varphi)\right],
\end{equation}
where $g^{(5)}$ is the metric and $R[g^{(5)}]$ the scalar curvature in the bulk;
$\varphi$ is a dilaton field living on the bulk and $V(\varphi)$ is a general potential energy. Note that we are using $\kappa_{5}^{2}=8\pi G_{5}=2$.

The simplest line element of the brane, embedded in the 5D bulk spacetime can be written as \cite{br}:
\begin{eqnarray}
ds^{2}_{5}&=&g_{AB}dx^{A}dx^{B}\nonumber\\
&=&dw^{2}+e^{2A}(dx^{2}+dy^{2}+dz^{2}-dt^{2}),
\end{eqnarray}
where $A$ is the warp factor which depends only on the five-dimensional (5D) coordinate $w$. For the scalar field $\varphi$ with the potential $V(\varphi)$, the 5D energy-momentum tensor is given by
\begin{equation}
T_{AB}=\partial_{A}\varphi\partial_{B}\varphi-g_{AB}\left[\frac{1}{2}\partial_{C}\varphi \;\partial^{C}\varphi+V(\varphi)\right],
\end{equation}
where $g_{AB}$ and $\varphi$ depend only on $w$.

The 5D gravitational and scalar field equations take the following forms
\begin{eqnarray}
3A^{\prime\prime}+6{A^{\prime}}^{2}&=&-\kappa_{5}^{2}e^{-2A}T_{00}=-\kappa_{5}^{2}\left[\frac{1}{2}{\varphi^{\prime}}^{2}+V(\varphi)\right] ,
     \\
6{A^{\prime}}^{2}&=&\kappa_{5}^{2}T_{44}=\kappa_{5}^{2}\left[\frac{1}{2}{\varphi^{\prime}}^{2}-V(\varphi)\right] ,
     \\
\varphi^{\prime\prime}+4A^{\prime}\varphi^{\prime}&=&\frac{d
V(\varphi)}{d\varphi},
\end{eqnarray}
respectively, where the prime denotes a derivative with respect to $w$.

In order to obtain a first-order equation, we introduce an auxiliary function $W$ according to \cite{de,Sa,Af,CM,DA}, which demands:
\begin{eqnarray}
A^{\prime}&=&-\frac{1}{3}W(\varphi), \\
\varphi^{\prime}&=&\frac{1}{2}\frac{\partial
W(\varphi)}{\partial\varphi},
\end{eqnarray}
while $V(\varphi)$ takes the following form \cite{de,Sa,Af,CM,DA}:
\begin{equation}
V(\varphi)=\frac{1}{8}\left(\frac{\partial
W(\varphi)}{\partial\varphi}\right)^{2}-\frac{1}{3}W(\varphi)^{2}.
\end{equation}

The $T_{00}$ distribution on the bulk, which will be analysed in detail below, is given by \cite{blg}
\begin{equation}
T_{00}=e^{2A}\left[\frac{1}{2}\left(\frac{\partial\varphi}{\partial
w}\right)^{2}+V(\varphi)\right].
\end{equation}
It can also be shown that, for models with an infinitely thin brane and Dirac delta distributions, the energy density is equal to the cosmological constant of the bulk plus the energy density on the brane, i.e., $\varepsilon=\Lambda_{5}^{\pm}+k\delta(w)$.

Moreover, it may be instructive to calculate the geodesic equation along the fifth dimension in a thick brane, in order to investigate the particle motion near the brane \cite{JS}. As mentioned before, thick brane models considered in this paper do not have Dirac delta singularities which enable easier direct calculations. To this end, we start with the geodesic equation:
\begin{eqnarray}
\frac{d^{2}x^{0}}{d\tau^{2}}+\Gamma^{0}_{AB}\frac{dx^{A}}{d\tau}\frac{dx^{B}}{d\tau}&=&0   \qquad
\Rightarrow  \qquad  \frac{d}{d\tau}\left(-2e^{2A}\dot{t}\right)=0,  \nonumber\\
\frac{d^{2}x^{4}}{d\tau^{2}}+\Gamma^{4}_{AB}\frac{dx^{A}}{d\tau}\frac{dx^{B}}{d\tau}&=&0 \qquad
\Rightarrow \qquad \ddot{w}+A'e^{2A}\dot{t}^{2}=0 ,
\end{eqnarray}
which leads to
\textbf{\begin{equation}\label{geo}
\ddot{w}+c_{1}^{2}f(w)=0 ,
\end{equation}}
where $c_{1}$ is a constant of integration and the function $f(w)$ is defined as
\begin{equation}
f(w)=A'(w)e^{-2A(w)}.
\end{equation}

Equation (\ref{geo}) is a  second order differential equation for $w$ and its solution depends critically on whether $f(w)/w$ is positive or negative. For  positive values of $f(w)/w$ one obtains periodic (exponential) solutions, respectively. Note that the periodic (negative) motion indicates particle confinement near the brane, while the exponential solutions implies that the reference point is unstable. However, this may point to the possibility that $w=0$ is different from the localization of the brane. In a periodic situation, by introducing a new quantity
$F(w)=c_{1}^{2}A'(w)e^{-2A(w)}$, one can write the geodesic equation in the following form
\begin{equation}
\ddot{w}+F(w)=0.
\end{equation}
The equilibrium point $w_{0}$ satisfies $F(w_{0})$=0. On the other hand, by expanding $F(w)$ around $w_{0}$, we have
\begin{equation}
F(w)=F(w_{0})+F'(w_{0})(w-w_{0})+... \,,
\end{equation}
and the geodesic equation leads to
\begin{equation}
\ddot{w}+F'(w_{0})(w-w_{0})=0.
\end{equation}
Taking into account a change of variable $\tilde{w}=w-w_{0}$, the geodesic equation reduces to
\begin{eqnarray}\label{ge}
\ddot{\tilde{w}}+F'(w_{0})\tilde{w}=0, \qquad  {\rm or} \qquad \ddot{\tilde{w}}+\Omega^{2}\tilde{w}=0 \,,
\end{eqnarray}
where $\Omega=\sqrt{F'(w_{0})}$, provided that $F'(w_{0})\geq0$.

It is essential to emphasise that in the RS-II brane model, the $KK$ zero mode corresponds to the massless graviton and the massive modes form a continuum which result in a small correction to Newtonian gravity at large distances \cite{YZYX,FDCR}. Free particles are only affected by the gravitational field and not directly by the scalar field. Any field or particle which has a direct coupling with the scalar field, will be further affected by extra force from the scalar field.
Moreover, even an exponential potential like $e^{-\alpha w^{2}}$ reduces to a harmonic potential for small amplitude oscillations ($e^{-\alpha w^{2}}\approx1-\alpha w^{2}+O(w^{4})$).

In the following section, we explore several models for thick branes and we will employ the linearized geodesic equation (\ref{ge}) for each model.

\section{Soliton models for the brane}\label{22}

\subsection{Sine-Gordon-based models}\label{22a}

The sine-Gordon (SG) model is a well-known integrable model which
has found interesting applications in various disciplines
\cite{Ri,Pey}. In fact, single and multiple (topological) soliton solutions of this system are found analytically through different mathematical methods
\cite{Ri,Pey}. The self-interaction potential for this model reads
\begin{equation}
\tilde{V}(\varphi)=\frac{a}{b}\left[1-cos(b\varphi)\right]\,,
\end{equation}
where $a$ and $b$ are free parameters of the model.
When considered as the brane potential, however, this potential
should be modified to become consistent with the Einstein equations.

\begin{figure*}
\epsfxsize=15.8cm\centerline{\epsfbox{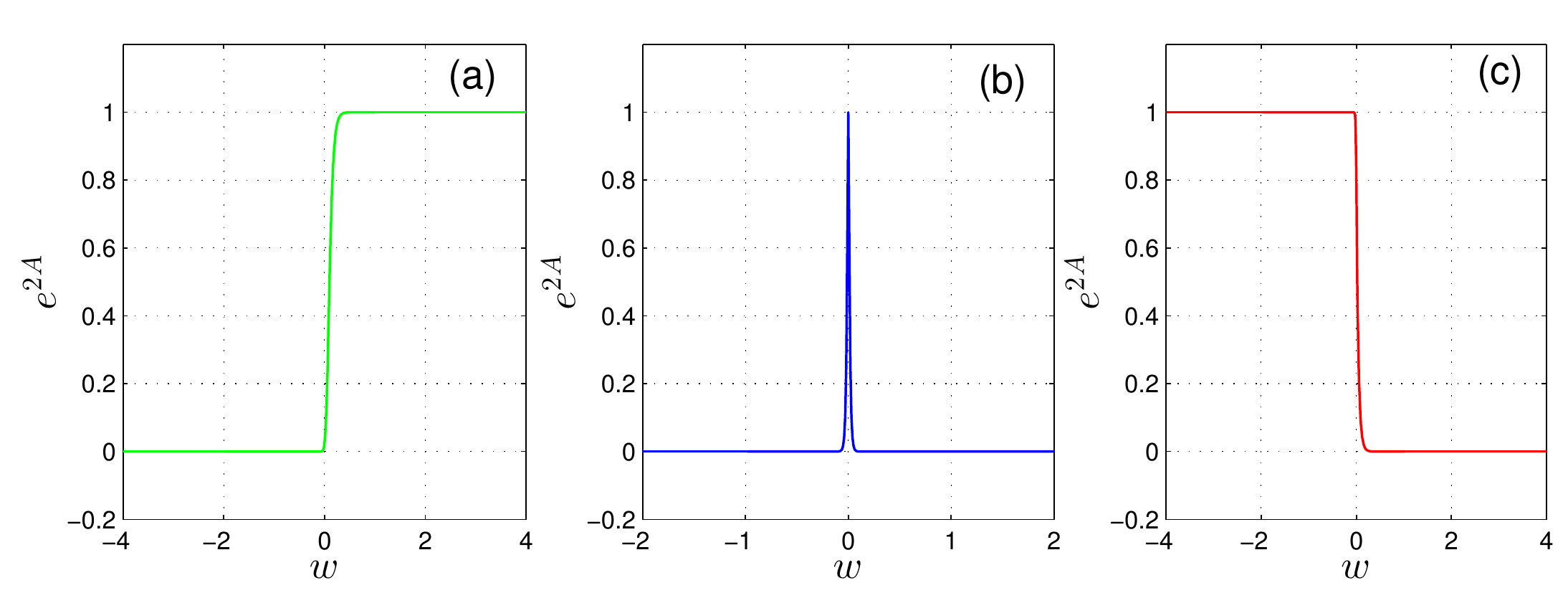}} \caption{The plots
depict the warp factor as a function of the fifth dimension for the
(a) SG with $a=100$ and $b=1$ , (b) $\varphi^4$ with $\alpha=1$ and
$\beta=100$ and (c) $\varphi^6$ with $\alpha=1$ and $\beta=100$
systems, respectively. For the SG and $\varphi^6$ systems there is an
asymmetry between the two sides of the brane and the $Z_2$ symmetry
is broken.\label{wp}}
\end{figure*}

The SG system has the following exact static kink solution \cite{Ri}:
\begin{equation}
\varphi(w)=\frac{4}{b}\arctan\left(e^{\sqrt{ab}w}\right),
  \label{SGdilaton}
\end{equation}
which is plotted in Fig. \ref{st}(a), for various values of parameters $a$ and $b$, which correspond to branes with different thicknesses. 
The formalism of our investigation is to keep the soliton solution of the flat space nonlinear equation and modify the scalar field potential in such a way that the soliton solution remains a solution of the full gravitating system. This is why the soliton solution remains the same. The form of the potential, however, changes accordingly. Taking into account the scalar field given by Eq. (\ref{SGdilaton}), and plugging it into the field equations, we obtain the following quantities
\begin{eqnarray}
W(\varphi)&=&-\frac{16\sqrt{ab}}{b^{2}}\left[\cos^{2}\left(\frac{b\varphi}{4}\right)\right],\nonumber\\
A&=& - \frac{8}{3b^{2}}\ln\left(\frac{1+e^{2\sqrt{ab}w}}{e^{2\sqrt{ab}w}}\right),\nonumber\\
\exp(2A)&=&\left(\frac{1+e^{2\sqrt{ab}w}}{e^{2\sqrt{ab}w}}\right)^{-\frac{16}{3b^{2}}},
\end{eqnarray}
where the warp factor is plotted in Fig. \ref{wp}(a). The corresponding modified potential for this model is given by
\begin{equation}
V(\varphi)=\frac{2a}{b}\sin^{2}\left(\frac{b\varphi}{2}\right)-\frac{64a}{3b^{3}}\left[1+\cos\left(\frac{b\varphi}{2}\right)\right]^{2}\,,
  \label{SGpotential}
\end{equation}
which is depicted in Fig. \ref{v}(a). Notice that this potential has two series of  non-degenerate vacuua, as in the DSG  (double sine-Gordon) system
potential \cite{Pey}. However, in the limit of $b\gg a$ these vacuua
tend to the same value (become degenerate), such as the potentials used in \cite{blg,CM}.

This system leads to the following $T_{00}$, from which the energy density can be obtained,
\begin{equation}
T_{00}=\left(\frac{1+e^{2\sqrt{ab}w}}{e^{2\sqrt{ab}w}}\right)^{\left(\frac{-16}{3b^{2}}\right)}\left[\frac{16}{3}\frac{a\left(3b^{2}e^{2\sqrt{ab}w}-16\right)}{b^{3}\left(1+e^{2\sqrt{ab}w}\right)^{2}}\right],
\end{equation}
which is plotted in Fig. \ref{de}(a). The energy momentum tensor is calculated according to:
\begin{equation}
T_{\mu\nu}=\partial_{\mu}\varphi\partial_{\nu}\varphi-g_{\mu\nu}\left[\frac{1}{2}\left(\partial_{\alpha}\varphi\right)\left(\partial^{\alpha}\varphi\right)+V(\varphi)\right],
\end{equation}
where the modified potential is used for $V$. The metric and the energy momentum tensor are checked to satisfy the Einstein equations. Moreover, note that the potential for any soliton model may be shifted by a constant, without affecting the soliton solutions. The minimum value of the potential (i.e., the classical vacuum), if negative, leads to  a negative energy density. This can be avoided by adding a positive constant to the self-interaction potential, when the scalar filed is  not coupled to gravity (i.e., in flat spacetime). In curved spacetime, however, this constant is non-trivial and plays the role of a cosmological constant of the bulk. Of course, a positive cosmological constant violates the strong energy condition.

Note that the energy density is localized at the brane and the thickness of the latter is given by
\begin{equation}
\triangle=\frac{1}{2\sqrt{ab}}.
\end{equation}
In this model, the Ricci and Kretschmann scalars are given by
\begin{eqnarray}
R=\frac{256}{9}\frac{a\left[-20+3\left(e^{2\sqrt{ab}w}\right)b^{2}\right]}{b^{3}\left[1+\left(e^{2\sqrt{ab}w}\right)\right]^{2}} \,,
\end{eqnarray}
and
\begin{eqnarray}
K=\frac{16384}{81}\frac{a^{2}\left[160-48\left(e^{2\sqrt{ab}w}\right)b^{2}+9\left(e^{4\sqrt{ab}w}\right)b^{4}\right]}{b^{6}\left[1+\left(e^{2\sqrt{ab}w}\right)\right]^{4}},
\end{eqnarray}
respectively. It can be seen that there is no singularity in the Ricci scalar
and/or Kretschmann scalar. In the limits of $w\rightarrow\pm\infty$, the Ricci scalar becomes
\begin{eqnarray}
\lim_{w\rightarrow+\infty} R&=&0, \\
\lim_{w\rightarrow-\infty} R&=&-\frac{5120}{9}\frac{a}{b^{3}},
\end{eqnarray}
and the limit of $w\rightarrow0$ yields
\begin{equation}
\lim_{w\rightarrow0} R=\frac{64}{9}\frac{(-20+3b^{2})a}{b^{3}}.
\end{equation}

Moreover, the mixed Einstein tensor components are given by:
\begin{eqnarray}
G^{0}_{0}&=&G^{1}_{1}=G^{2}_{2}=G^{3}_{3}
=-\frac{32}{3}\frac{a\left(-16+3e^{2\sqrt{ab}w}b^{2}\right)}{b^{3}\left(1+e^{2\sqrt{ab}w}\right)^{2}},
\end{eqnarray}
\begin{eqnarray}
G^{4}_{4}&=&\frac{512}{3}\frac{a}{b^{3}\left(1+e^{2\sqrt{ab}w}\right)^{2}},
\end{eqnarray}
respectively. Note that all the components of the Einstein tensor in the limits
$w\rightarrow\pm\infty$ become
\begin{eqnarray}
\lim_{w\rightarrow+\infty} G^{A}_{B}&=&0,
      \\
\lim_{w\rightarrow-\infty} G^{A}_{B}&=&\frac{512}{3}\frac{a}{b^{3}},
\end{eqnarray}
However, in the limit of $w\rightarrow0$ (for $\mu=\nu=0,1,2,3$) the
Einstein tensor is given by
\begin{equation}
\lim_{w\rightarrow0}G^{\mu}_{\nu}=\frac{8}{3}\frac{(16-3b^{2})a}{b^{3}}\,\delta^{\mu}_{\nu}\,,
\end{equation}
and one can interpret it as the cosmological constant on the brane, i.e.,
$G^{i}_{j}\propto\Lambda\delta^{i}_{j}$, with $\Lambda=\frac{8}{3}\frac{(16-3b^{2})a}{b^{3}}$.

These results can be interpreted in that we have a broken $Z_{2}$-symmetry in the bulk, as the two sides of the brane differ completely. On the right ($w\rightarrow+\infty$), the Einstein tensor and consequently the cosmological constant of the bulk vanish, so the bulk is asymptotically Minkowski. However, on the other side of the brane, these quantities are nonzero and equal to the constant value
$512 a/(3b^{3})$, and as a result the bulk would be
de Sitter. The Ricci scalar and the Einstein tensor component $G^{0}_{0}$ are plotted Fig. \ref{r}(a) and Fig. \ref{gw}(a), respectively.

Furthermore, by calculating the field equations,
\begin{equation}
G_{AB}=\kappa_{5}^{2}T_{AB}.
\end{equation}
one verifies that $\kappa_{5}^{2}=2$, which is consistent with
the usual normalization notation \cite{Sa,CM,KS}.
As pointed out in DeWolfe {\it et al}. \cite{de}, in the stiff limit where $ab\rightarrow\infty$, the wall
reduces to the step function and the energy density approaches a
$\delta$-function (see Figs. \ref{st}(a) and \ref{de}(a)).

The local confining gravitational field of the brane is best observed by
looking at the geodesic equation of a test particle moving only in
the direction of the extra dimension, as explained in the previous section:
\begin{equation}
\ddot{w}+c_{1}^{2}\frac{8}{9}\frac{32^{\frac{1}{b^{2}}}2^{\frac{1}{3b^{2}}}a\left(3b^{2}+16\right)}{b^{3}}w\approx
c_{1}^{2}\frac{8}{3}\frac{32^{\frac{1}{b^{2}}}2^{\frac{1}{3b^{2}}}\sqrt{ab}}{b^{2}}.
\end{equation}
This proves the confining effect of the scalar field, and $c_{1}$ is an integration constant [see Eq. \ref{geo}]. For small amplitude oscillations, the relativistic motion reduces to a classical motion in a Newtonian classical potential. Since we have considered the potential up to second order, relativistic effects can be ignored and the corresponding quantum energy levels are therefore those of a non-relativistic quantum particle. So, if interpreted as a quantum oscillator, one can assign an energy to each quantum state given by $E_n=(n+1/2)\hbar\omega$, where
\begin{equation}
\hbar
\omega=\Omega=\sqrt{F'(w_{0})}\approx
c_{1}\sqrt{\frac{8}{9}\frac{32^{\frac{1}{b^{2}}}2^{\frac{1}{3b^{2}}}a(3b^{2}+16)}{b^{3}}}.
\end{equation}

However, an important point is in order. In what follows, we explore the confining effect of the brane, only up to second order in the potential. Even if the classical test particle is confined up to this order, large amplitude oscillations will involve nonlinear effects and this might exploit confinement. Quantum mechanically, the full nonlinear potential might lead to tunneling and thus de-confinement. It is well known that massive $KK$ modes are not confined to the brane.

\subsection{$\varphi^{4}$-based model}

The $\varphi^4$ models are well known for having simple soliton-like solutions, although it is not strictly integrable like the SG system. This model is also the central ingredient in the Goldtone and Higgs mechanisms. Spontaneous breakdown of the $Z_2$ symmetry in the complex version of the
$\varphi^4$ model leads to the appearance of the Goldstone mode and once coupled with a Gauge field, it causes the Gauge boson to acquire mass \cite{gui}. This model is therefore frequently used for building thick branes.

\begin{figure*}
\epsfxsize=16cm\centerline{\epsfbox{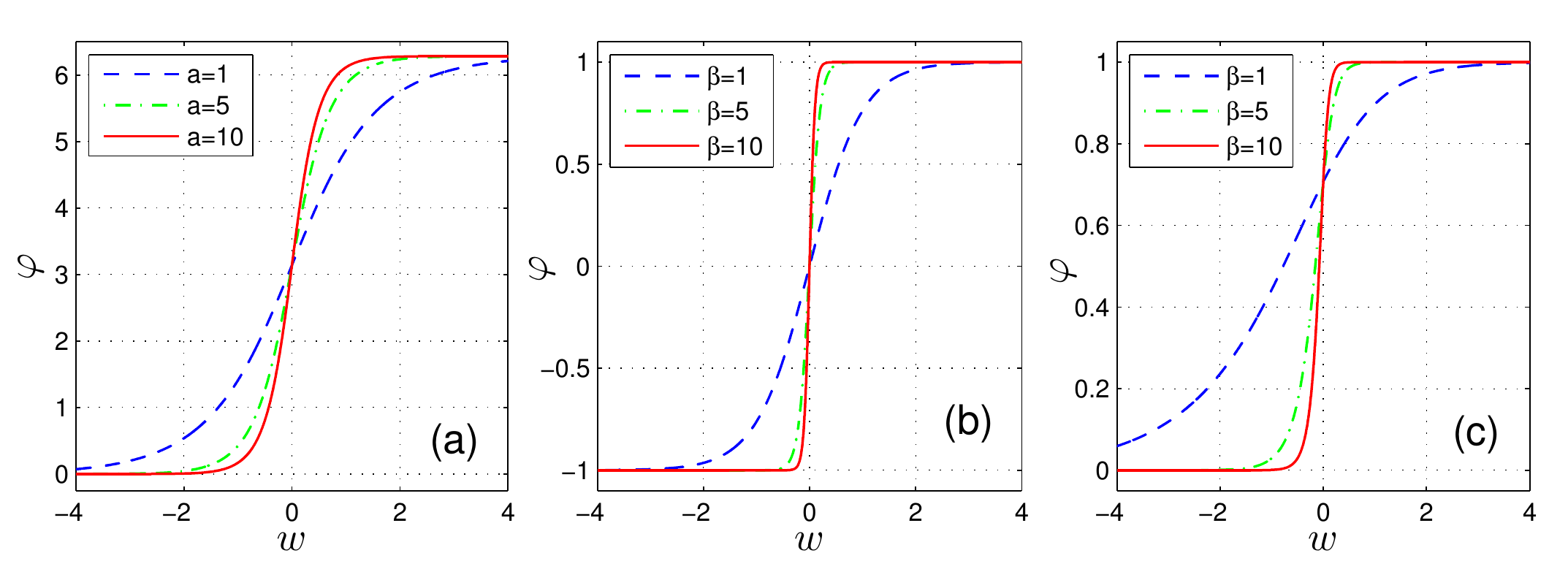}} \caption{Soliton
solutions as a function of the five-dimensional coordinate $w$ for the following models: (a) SG for $b=1$, (b) $\varphi^4$ for $\alpha=1$ and (c) $\varphi^6$ for $\alpha=1$ systems.
Dashed, dotted-dashed, and continuous curves correspond to solitons
with decreasing brane thickness. In the limit of an infinite $a/\beta$
parameter, the soliton approaches the step function. \label{st}}
\end{figure*}

\begin{figure*}
\epsfxsize=15.8cm\centerline{\epsfbox{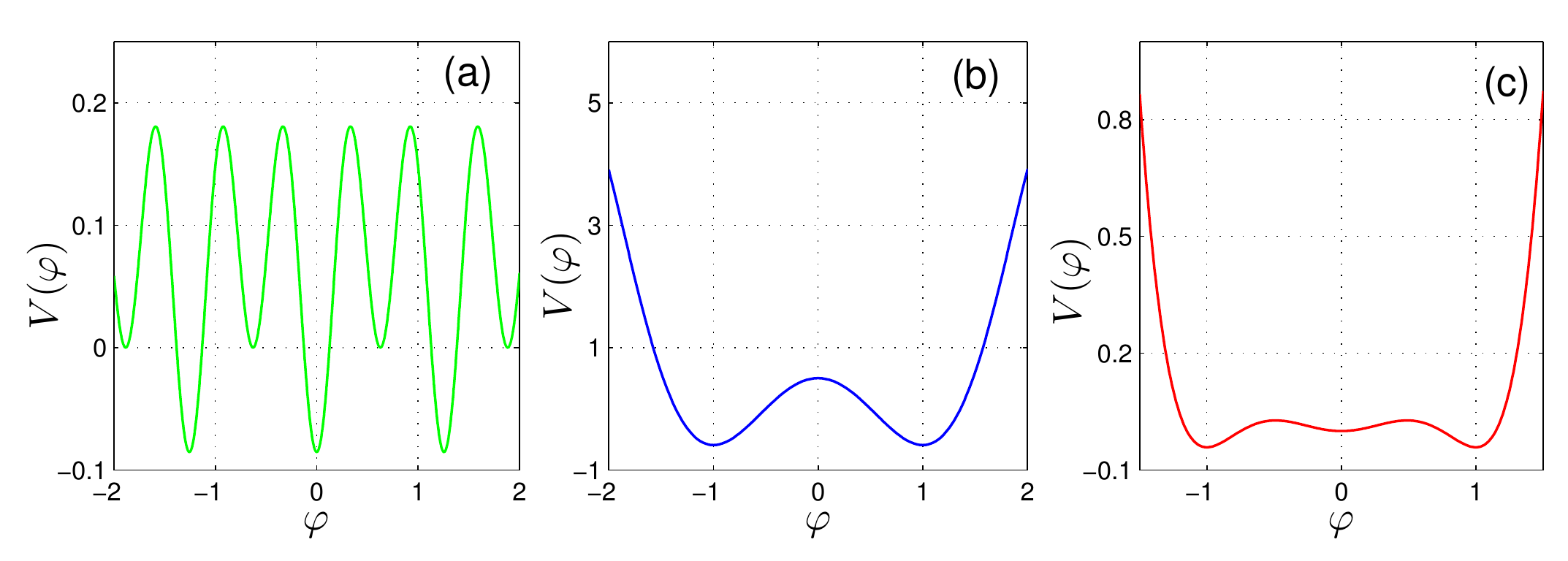}} \caption{The plots depict
the modified soliton potential as a function of the scalar field
for (a) SG with $a=b=10$, (b) $\varphi^4$ with $\alpha=\beta=1$ and (c) $\varphi^6$ with
$\alpha=\beta=1$ systems.  The potential of the SG and $\varphi^6$
systems have non-degenerate vacuua. In contrast, the $\varphi^4$
potential has degenerate vacuua and this leads to a stable,
topological solitonic brane. For $\varphi^{4}$ and $\varphi^{6}$ systems the number of extrema in the Figure is only a result of the range of the plot. In a wider range plot, other extrema appear; these may render the system globally unstable. We have only claimed local stability.\label{v}}
\end{figure*}

For the $\varphi^{4}$-based model, we have \cite{blg}
\begin{equation}
\tilde{V}(\varphi)=\frac{\beta^{2}}{2\alpha^{2}}\left(\varphi^{2}-\alpha^{2}\right)^{2},
\end{equation}
where $\alpha$ and $\beta$ are constants. The kink solution reads
\begin{equation}
\varphi(w)=\alpha\tanh(\beta w),
\end{equation}
which is depicted in Fig. \ref{st}(b). Proceeding in a similar manner as in the previous case, we have the following solutions:
\begin{eqnarray}
W(\varphi)&=&\frac{2}{3}\frac{\beta\varphi\left(3\alpha^{2}-\varphi^{2}\right)}{\alpha},
     \\
A&=&-\frac{4}{9}\alpha^{2}\ln\left[\cosh(\beta w)\right]+\frac{1}{9}\alpha^{2}\frac{1}{\cosh^{2}(\beta w)},
      \\
\exp(2A)&=&\left[\cosh(\beta w)\right]^{-\frac{8}{9}\alpha^{2}}\exp\left[\frac{2}{9}\alpha^{2}\frac{1}{\cosh^{2}(\beta w)}\right],
\end{eqnarray}
respectively, and the potential is obtained as
\begin{equation}
V(\varphi)=\frac{1}{2}\alpha^{2}\beta^{2}\left(1-\frac{\varphi^{2}}{\alpha^{2}}\right)^{2}-\frac{4}{27}\varphi^{2}\alpha^{2}\beta^{2}\left(3-\frac{\varphi^{2}}{\alpha^{2}}\right)^{2}\,,
\end{equation}
which is plotted in Fig. \ref{v}(b).
It is seen that while $\tilde{V}(\varphi)$ was $O(\varphi^{4})$,
$V(\varphi)$ is $O(\varphi^{6})$.

\begin{figure*}
\epsfxsize=15.8cm\centerline{\epsfbox{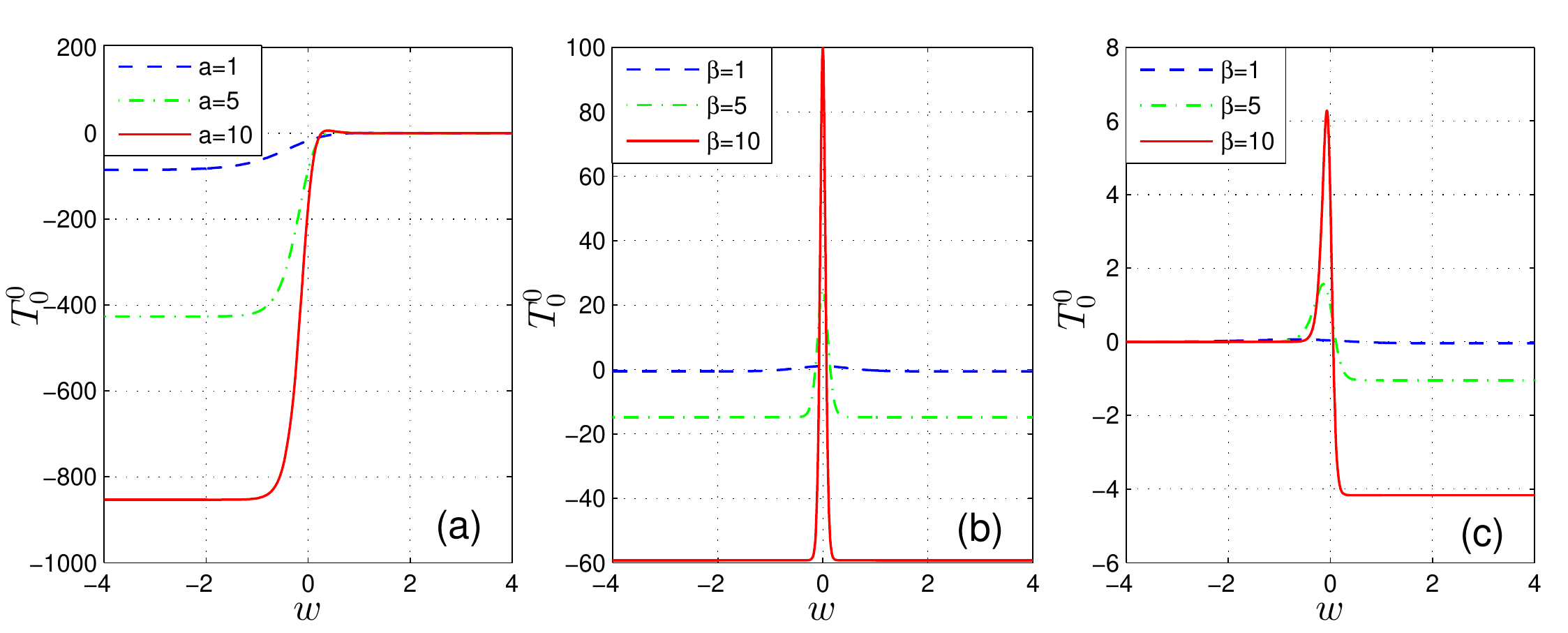}} \caption{$T^{0}_{0}$ as
a function of the fifth dimension coordinate $w$ for the following systems:
(a) SG with $b=1$, (b) $\varphi^4$ with $\alpha=1$ and (c) $\varphi^6$ with $\alpha=1$. The dashed, dotted-dashed and continuous curves correspond to
increasingly thin branes. The brane becomes infinitely thin (delta
function) as the parameter $a/\beta$ tends to infinity.\label{de}}
\end{figure*}

The corresponding $T_{00}$ is given by
\begin{eqnarray}
T_{00}=-\frac{1}{27}\frac{\alpha^{2}\beta^{2}\exp \left[-\frac{2}{9}\alpha^{2}\tanh^{2}(\beta w)\right]}{\cosh^{\left(6-\frac{8}{9}\alpha^{2}\right)}(\beta w)}
  \Big[ -4\alpha^{2} -27\cosh^{2}(\beta w)
   \nonumber  \\
+16\alpha^{2}\cosh^{6}(\beta
w)-12\alpha^{2}\cosh^{2}(\beta w)\Big]\,,
\end{eqnarray}
which is depicted in Fig. \ref{de}(b), where one verifies that the energy density is localized at the brane. The brane thickness becomes $\triangle=\beta^{-1}$.

\begin{figure*}
\epsfxsize=16cm\centerline{\epsfbox{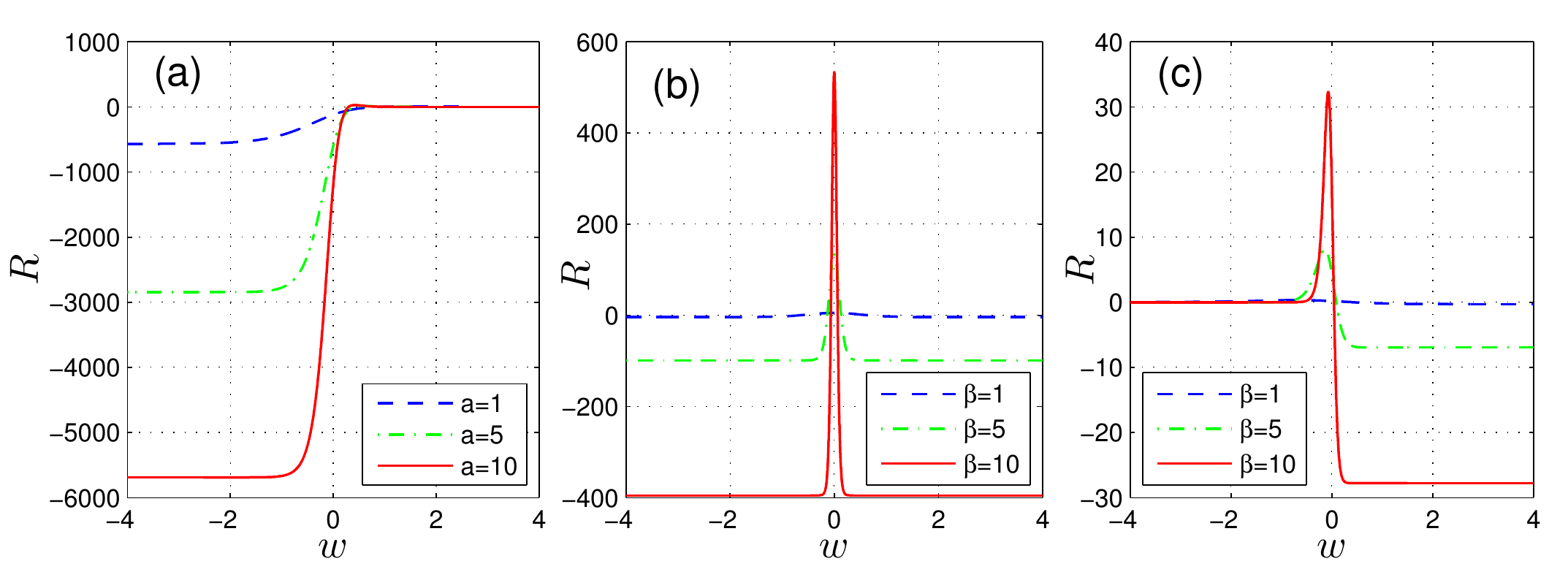}} \caption{Ricci scalar as
a function of the fifth dimension coordinate $w$ for the following systems: (a) SG
with $b=1$, (b) $\varphi^4$ with $\alpha=1$  and (c) $\varphi^6$ with $\alpha=1$. The Ricci
scalar is different on the two sides of the brane, due to the effect of the warp factor. \label{r}}
\end{figure*}

\begin{figure*}
\epsfxsize=15.8cm\centerline{\epsfbox{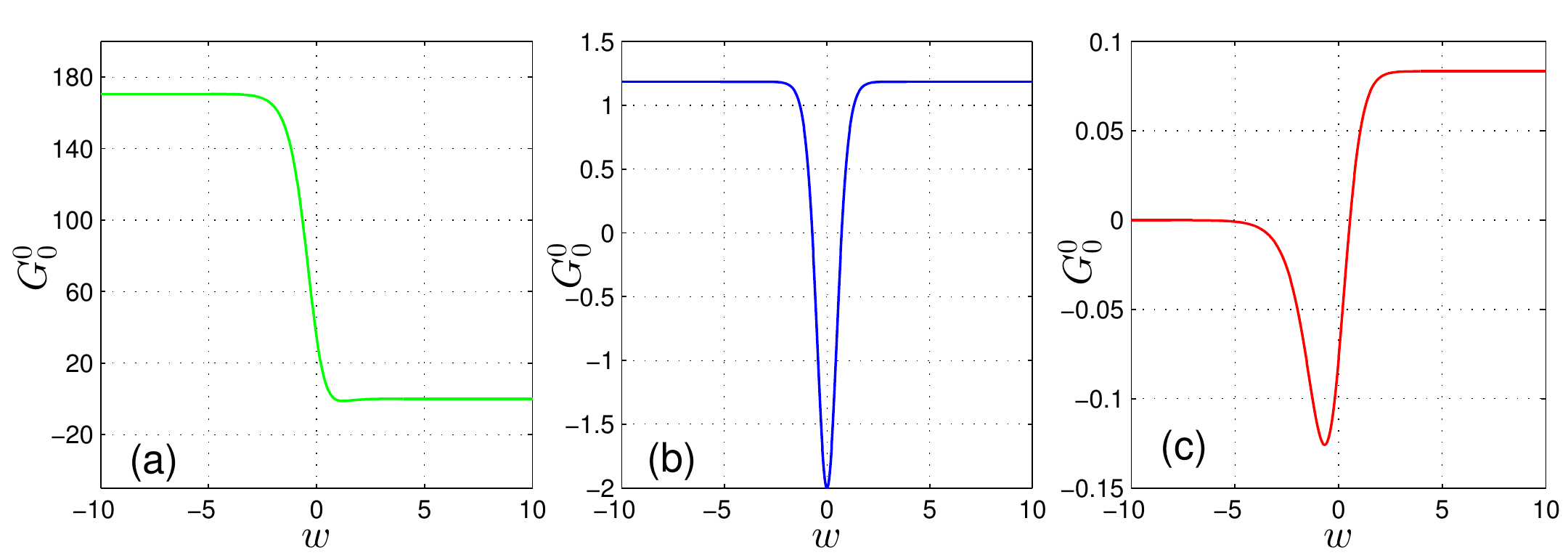}} \caption{The plots
depict the Einstein tensor component $G_{0}^{0}$ for the (a) SG with
$a=b=1$, (b) $\varphi^4$ with $\alpha=\beta=1$ and (c) $\varphi^6$ with
$\alpha=\beta=1$ systems. Note that this quantity approaches
different constant values for the SG and $\varphi^{6}$ systems,
while the $\varphi^{4}$ system is $Z_{2}$-symmetric.\label{gw}}
\end{figure*}

Moreover, one can show that the Ricci and Kretschmann scalars are given by
\begin{eqnarray}
R=-\frac{16}{81}\frac{\alpha^{2}\beta^{2}}{\cosh^{6}(\beta w)}
\Big[-15\alpha^{2}\cosh^{2}(\beta
w)-5\alpha^{2}
    +20\alpha^{2}\cosh^{6}(\beta w)-27\cosh^{2}(\beta
w)\Big],
\end{eqnarray}
which is depicted in Fig. \ref{r}(b), and
\begin{eqnarray}
K&=&\frac{64}{6561}\frac{\alpha^{4}\beta^{4}}{\cosh^{12}(\beta w)}\Big[90\alpha^{4}\cosh^{4}(\beta w)
-80\alpha^{4}\cosh^{6}(\beta w)
-240\alpha^{4}\cosh^{8}(\beta w)
  \nonumber   \\
&&+160\alpha^{4}\cosh^{12}(\beta w)+60\alpha^{4}\cosh^{2}(\beta w)
 +10\alpha^{4}
+324\alpha^{2}\cosh^{4}(\beta w)
   \nonumber  \\
&&+108\alpha^{2}\cosh^{2}(\beta w)-432\alpha^{2}\cosh^{8}(\beta w)+729\cosh^{4}(\beta w)\Big],
\end{eqnarray}
respectively. The mixed Einstein tensor components take the following form
%
\begin{eqnarray}
G^{0}_{0}=G^{1}_{1}=G^{2}_{2}= G^{3}_{3} =\frac{2}{27}
     \frac{\alpha^{2}\beta^{2}}{\cosh^{6}(\beta w)}
\big[-4\alpha^{2}-12\alpha^{2}\cosh^{2}(\beta w)
    \nonumber \\
+16\alpha^{2}\cosh^{6}(\beta w)-27\cosh^{2}(\beta w)\big],
\end{eqnarray}
%
%
\begin{eqnarray}
G^{4}_{4}&=&\frac{8}{27}\frac{\alpha^{4}\beta^{2}\left[-3\cosh^{2}(\beta w)-1+4\cosh^{6}(\beta w)\right]}{\cosh^{6}(\beta w)}.
\end{eqnarray}
where the $G^0_0$ component is depicted in Fig. \ref{gw}(b).

As for the previous SG model, we determine the limits $w\rightarrow\pm\infty$ for all the components of the Einstein tensor components, which are given by:
\begin{eqnarray}
\lim_{w\rightarrow+\infty} G^{A}_{B}&=&\frac{32}{27}\alpha^{4}\beta^{2},
      \\
\lim_{w\rightarrow-\infty}G^{A}_{B}&=&\frac{32}{27}\alpha^{4}\beta^{2},
\end{eqnarray}
and in the limit of $w\rightarrow0$ the Einstein tensor components (for
$\mu=\nu=0,1,2,3$) takes the form
\begin{equation}
\lim_{w\rightarrow0}
G^{\mu}_{\nu}=-2\alpha^{2}\beta^{2}\delta^{\mu}_{\nu}\,.
\end{equation}
Note that in this model the cosmological constant on the brane $\Lambda$ would be $-2\alpha^{2}\beta^{2}$. Taking into account all of the above considerations, we verify that the Einstein equations are given consistently by $G_{AB}=\kappa_{5}^{2}T_{AB}$
where $\kappa_{5}^{2}=2$.

The geodesic equation for a test particle moving in the direction of
the fifth dimension one obtains
\begin{equation}
\ddot{w}+c_{1}^{2}\frac{2}{3}\alpha^{2}\beta^{2}w=0\,,
\end{equation}
which corresponds to a linearized quantum mode of energy
$\hbar\omega=\Omega=\sqrt{F'(w_{0})}=\sqrt{\frac{2}{3}}c_{1}\alpha\beta$.

It is seen that the resulting potential for the $\varphi^{4}$ model is an odd function with respect to $w$, while the $A$ function is even.  This property is not verified in the SG and $\varphi^{6}$ models, where the latter is discussed below. This results in an asymmetry in the corresponding properties (such as the energy density). However, by an appropriate selection of the model parameters, one can restore the $Z_2$ symmetry in the $\varphi^4$ and $\varphi^6$ cases, which is commonly used in brane models.

\subsection{$\varphi^{6}$-based model}

For this model, we have the following potential:
\begin{equation}
\tilde{V}(\varphi)=\frac{\beta^{2}}{4\alpha^{2}}\varphi^{2}\left(\varphi^{2}-\alpha^{2}\right)^{2}\,,
\end{equation}
and as a result, the kink solution is given by \cite{hosein}
\begin{equation}
\varphi (w)=\frac{\alpha}{\sqrt{1+e^{(-\sqrt{2}\alpha\beta w)}}},
\end{equation}
where $\alpha$ and $\beta$ are constant (as in the
$\varphi^{4}$ model). The kink solution is depicted in Fig. \ref{st}(c).
For this model the potential $W$, and the warp factor $A$ are given by the following expressions
\begin{eqnarray}
W(\varphi)&=&\frac{\sqrt{2}}{4}\frac{\beta\varphi^{2}(2\alpha^{2}-\varphi^{2})}{\alpha},
      \\
A&=&-\frac{\alpha^{2}}{12}\left[\frac{1}{1+e^{-\sqrt{2}\alpha\beta w}}+\ln\left(\frac{1+e^{-\sqrt{2}\alpha\beta w}}{e^{-\sqrt{2}\alpha\beta w}}\right)\right],
      \\
\exp(2A)&=&\left(\frac{e^{-w\sqrt{2}\beta\alpha}}{1+e^{-w\sqrt{2}\beta\alpha}}\right)^{\frac{\alpha^{2}}{6}}e^{-\frac{1}{6}\frac{\alpha^{2}}{1+e^{-w\sqrt{2}\beta\alpha}}}\,,
\end{eqnarray}
respectively. Thus, for the $\varphi^{6}$ system the self-interaction potential takes the form
\begin{equation}
V(\varphi)=\frac{1}{4}\frac{\beta^{2}\varphi^{2}\left(\alpha^{2}-\varphi^{2}\right)^{2}}{\alpha^{2}}-\frac{1}{24}\frac{\beta^{2}\varphi^{4}\left(2\alpha^{2}-\varphi^{2}\right)^{2}}{\alpha^{2}}\,,
\end{equation}
which is depicted in Fig. \ref{v}(c). Note that $V(\varphi)$ has raised to $O(\varphi^{8})$.

The energy-momentum tensor component $T_{00}$ is given by:
\begin{eqnarray}
T_{00}&=& -\frac{1}{24}\left(\frac{e^{-w\sqrt{2}\beta\alpha}}{1+e^{-w\sqrt{2}\beta\alpha}}\right)^{\frac{\alpha^{2}}{6}}
\frac{e^{-\frac{1}{6}\frac{\alpha^{2}}{1+e^{-w\sqrt{2}\beta\alpha}}}}{\left(1+e^{-w\sqrt{2}\beta\alpha}\right)^{4}}  \Big[\alpha^{4}\beta^{2}\Big(-12e^{-2w\sqrt{2}\beta\alpha}
 \nonumber   \\
&&-12e^{-3w\sqrt{2}\beta\alpha}+4\alpha^{2}e^{-w\sqrt{2}\beta\alpha}
+\alpha^{2}+4\alpha^{2}e^{-2w\sqrt{2}\beta\alpha}\Big)\Big],
\end{eqnarray}
which is plotted in Fig. \ref{de}(c). The thickness of this brane is given by $\triangle=(\sqrt{2}\alpha\beta)^{-1}$.

For this system the Ricci scalar is given by
\begin{eqnarray}
R= -\frac{1}{18}
\frac{\alpha^{4}\beta^{2}e^{w\sqrt{2}\beta\alpha}}{\left(e^{w\sqrt{2}\beta\alpha}+1\right)^{4}}
\Big[20\alpha^{2}e^{w\sqrt{2}\beta\alpha}-48e^{w\sqrt{2}\beta\alpha}
   \nonumber \\
   + 5\alpha^{2}\left(e^{3w\sqrt{2}\beta\alpha}\right)
+20\alpha^{2}\left(e^{2w\sqrt{2}\beta\alpha}\right)-48 \Big],
\end{eqnarray}
which is plotted in Figs. \ref{r}(c), and the Kretschmann scalar by
\begin{eqnarray}
K =\frac{1}{648}\alpha^{8}\beta^{4}\frac{e^{2w\sqrt{2}\beta\alpha}}{\left(e^{w\sqrt{2}\beta\alpha}+1\right)^{8}}
\Bigg[1152+5\alpha^{4}\left(e^{6w\sqrt{2}\beta\alpha}\right)
+40\alpha^{4}\left(e^{5w\sqrt{2}\beta\alpha}\right)^{5}
   \nonumber  \\
+120\alpha^{4}\left(e^{4w\sqrt{2}\beta\alpha}\right)
+160\alpha^{4}\left(e^{3w\sqrt{2}\beta\alpha}\right)
+80\alpha^{4}\left(e^{2w\sqrt{2}\beta\alpha}\right)
    \nonumber  \\
-480\alpha^{2}\left(e^{3w\sqrt{2}\beta\alpha}\right)
+1152\left(e^{2w\sqrt{2}\beta\alpha}\right)
-768\alpha^{2}\left(e^{2w\sqrt{2}\beta\alpha}\right)
   \nonumber \\
-384\alpha^{2}e^{w\sqrt{2}\beta\alpha}
-96\alpha^{2}\left(e^{4w\sqrt{2}\beta\alpha}\right)+2304e^{w\sqrt{2}\beta\alpha}\Big].
\end{eqnarray}
The mixed Einstein tensor components are given by
\begin{eqnarray}
G^{0}_{0}=G^{1}_{1}=G^{2}_{2}=G^{3}_{3} =\frac{\alpha^{4}\beta^{2}e^{w\sqrt{2}\beta\alpha}}{12\left(e^{w\sqrt{2}\beta\alpha}+1\right)^{4}}
\Big(\alpha^{2}e^{3w\sqrt{2}\beta\alpha}
   \nonumber  \\
+4\alpha^{2}e^{2w\sqrt{2}\beta\alpha}+4\alpha^{2}e^{w\sqrt{2}\beta\alpha}-12e^{w\sqrt{2}\beta\alpha}-12\Big),
\end{eqnarray}
\begin{eqnarray}
G^{4}_{4}&=&\frac{1}{12}\frac{\alpha^{6}\beta^{2}e^{2w\sqrt{2}\beta\alpha}\left(e^{2w\sqrt{2}\beta\alpha}+4e^{w\sqrt{2}\beta\alpha}+4\right)}{\left(e^{w\sqrt{2}\beta\alpha}+1\right)^{4}}\,,
\end{eqnarray}
where the $G^0_0$ component is depicted in Fig. \ref {gw}(c).
These tensor components reduce to the following in the limit of
$w\longrightarrow\pm\infty$:
\begin{eqnarray}
\lim_{w\rightarrow+\infty} G^{A}_{B}&=&\frac{1}{12}\alpha^{6}\beta^{2},\nonumber\\
\lim_{w\rightarrow-\infty}G^{A}_{B}&=&0,
\end{eqnarray}
and in the limit of $w\rightarrow0$, the Einstein tensor is (for
$\mu=\nu=0,1,2,3$):
\begin{equation}
\lim_{w\rightarrow0}
G^{\mu}_{\nu}= \frac{1}{8}\alpha^{4}\beta^{2} \left(\frac{3}{8}\alpha^{2}-1\right)\, \delta^{\mu}_{\nu}\,.
\end{equation}

As in the previous cases, the Einstein equation in the bulk $G_{AB}=\kappa_{5}^{2}T_{AB}$ is found self-consistently
with $\kappa_{5}^{2}=2$, and which leads to the following  geodesic
equation
\begin{eqnarray}
\ddot{w}+c_{1}^{2}\frac{1}{192}\exp\left[\frac{\alpha^{2}}{12}\left(2\ln2+1\right)\right]\alpha^{4}\beta^{2}\left(8+3\alpha^{2}\right)w
   \nonumber  \\
\approx
\frac{1}{16}\exp\left[\frac{1}{12}\alpha^{2}\left(2\ln2+1\right)\right]\alpha^{3}\beta\sqrt{2}\,.
\end{eqnarray}
The quantum mode energy is thus given by
\begin{eqnarray}
\hbar \omega &=& \Omega=\sqrt{F'(w_{0})}
   \nonumber  \\
& \approx &
c_{1}\frac{\alpha^{2}\beta}{24}\sqrt{3}\sqrt{\exp\left(\frac{\alpha^{2}}{12}\left(2\ln(2)+1\right)\right)\left(8+3\alpha^{2}\right)}\,.
\end{eqnarray}

\section{Brane stability}\label{33}

In this section, we examine small perturbations about the soliton branes obtained
in the previous sections. To this end, the metric and the scalar
field are perturbed about the static brane and the resulting
equations of motion are linearized in the proper gauge
\cite{de,Sa,Af,CM,DA}.  The linearized equation turns out to be a
Schr\"{o}dinger-like equation with a potential $U(z)$ which determines
the linear modes. Unfortunately, for the models considered, this potential is too complicated to be handled analytically, or
to be used for finding the corresponding modes. Therefore, we will
only examine the potential near its minimum up to second order in
$z$.

In order to check the stability of the brane models, we use the standard scalar-tensor-vector (STV) decompositions \cite{YZYX}. For this purpose, a new variable $z$ is considered in such a way that $dz=e^{-A(w)} dw$, so the metric can be considered as \cite{YZYX}:
\begin{equation}
g_{MN}=a^2(z)\eta_{MN},
\end{equation}
where $a(z)=e^{A(z)}$. On the other hand, one can linearize the action of the system given by Eq.(\ref{ac}) along with the metric, via $\delta\varphi$ and $\delta g_{MN}\equiv a^{2}(z)h_{MN}$ which is given by \cite{YZYX}:
\begin{eqnarray}
\delta^{2}S &=&\frac{1}{2}\int d^{5}x a^{3}\Big[\partial_{M}h_{NP}\partial^{P}h^{MN}-\partial^{M}h\partial^{N}h_{MN}
 \nonumber \\
&&- \frac{1}{2}\partial_{P}h_{MN}\partial^{P}h^{MN}+\frac{1}{2}\partial^{P}h\partial_{p}h+3\frac{a'}{a}\left(h\partial^{\mu}h_{\mu z}-h_{zz} h'\right)
 \nonumber \\
&&+2\kappa_{5}^{2}\left(a^{2}V_{\varphi\varphi}\left(\delta\varphi\right)^{2}+2h^{Mz}\varphi '\partial_{M}\delta\varphi+\varphi'h' \delta\varphi-
\partial^{M}\delta\varphi\partial_{M}\delta\varphi\right)\Big].
\end{eqnarray}
where $\partial^{M}=\eta^{MN}\partial_{N}$, $\partial^{\mu}=\eta^{\mu\nu}\partial_{\nu}$, $h=\eta^{MN}h_{MN}$, and the prime denotes a derivative with respect to $z$. Moreover, by introducing the following vector and tensor perturbations \cite{YZYX,51}:
\begin{eqnarray}
h_{\mu z}&=&\partial_{\mu}F+G_{\mu}, \nonumber\\
h_{\mu\nu}&=&\eta_{\mu\nu}\varphi+\partial_{\mu}\partial_{\nu}B+\partial_{\mu}C_{\nu}+\partial_{\nu}C_{\mu}+D_{\mu\nu},
\end{eqnarray}
where $C_{\mu}$, $G_{\mu}$ and $D_{\mu\nu}$ are the transverse vector and tensor perturbations, respectively, one obtains
\begin{eqnarray}
\partial^{\mu}C_{\mu}&=0=&\partial^{\mu}G_{\mu}, \nonumber\\
\partial^{\mu}D_{\mu\nu}&=0=&D_{\mu}^{\mu}.
\end{eqnarray}

By using this STV decomposition method, one can decompose $\delta^{(2)}S$ into the following decoupled parts \cite{YZYX,51}:
\begin{eqnarray}
\delta^{(2)}S_{vector}&=&\frac{1}{2}\int d^{5}x\hat{v}^{\mu}\Box^{(4)}\hat{v}_{\mu},\nonumber\\
\delta^{(2)}S_{tensor}&=&\frac{1}{4}\int d^{5}x\hat{D}^{\mu\nu}\left[\Box^{(4)}\hat{D}_{\mu\nu}+\hat{D}_{\mu\nu}''-\frac{(a^{\frac{3}{2}})''}{a^{\frac{3}{2}}}\hat{D}^{\mu\nu}\right],
\end{eqnarray}
where
\begin{equation}
\hat{v}^{\mu}=a^{\frac{3}{2}}(G_{\mu}-C_{\mu}'),  \qquad  \hat{D}^{\mu\nu}=a^{\frac{3}{2}}D^{\mu\nu}, \qquad   \Box^{(4)}=\partial^{\mu}\partial_{\mu}.
\end{equation}
The scalar perturbations of the action, in turn, lead to two parts \cite{YZYX,51}:
\begin{equation}\label{s1}
\delta^{(2)}S_{scalar-1}=\int d^{5}x a^{3}\Big\{3\frac{a'}{a}h_{zz}-2\varphi'-2\kappa_{5}^{2}{\cal L}_{X}\varphi'\delta\varphi \Big\}\Box^{(4)}\psi,
\end{equation}
with $\psi=F-\frac{1}{2}B'$ and
\begin{eqnarray}\label{s2}
\delta^{(2)}S_{scalar-2}&=&\frac{1}{2}\int d^{5}x a^{3} \Big[-3\varphi\Box^{(4)}\varphi-3h_{zz}\Box^{(4)}\varphi+6\varphi'\varphi'\nonumber\\
&&-3\frac{a'}{a}h_{zz}(h_{zz}'+4\varphi')+2\kappa_{5}^{2}\left(\delta\varphi\Box^{(4)}\delta\varphi+a^{2}{\cal L}_{\varphi\varphi}(\delta\varphi)^{2}\right.\nonumber\\
&&\left.+2 h_{zz}\varphi'\delta\varphi'+\varphi'(h_{zz}'+4\varphi')\delta\varphi-(\delta\varphi')^{2}\right)\Big].
\end{eqnarray}
After eliminating $h_{zz}$ in action (\ref{s2}), by taking into account Eq.(\ref{s1}), and doing some simplifications, one obtains \cite{YZYX,51}:
\begin{equation}
\delta^{(2)}S_{scalar-2}=\int d^{5}x \hat{\mathcal{G}}\left\{\Box^{(4)}\hat{\mathcal{G}}+\hat{\mathcal{G}}''-\frac{\theta''}{\theta}\hat{\mathcal{G}}\right\},
\end{equation}
where $\mathcal{G}$ and $\theta$ are gauge invariant variables and specific functions of $a$, and are given by \cite{YZYX,51}:
\begin{eqnarray}
\mathcal{G}&=&\frac{\kappa_{5}^{2}}{2}a^{\frac{3}{2}}\left(2\delta\varphi-\frac{\varphi' a}{a'\varphi}\right),\nonumber\\
\theta &=& a^{\frac{3}{2}}\frac{\varphi' a}{a'},
\end{eqnarray}
respectively. Finally, the equation of normal modes of the linear perturbations are:
\begin{eqnarray}
\text{vector:} &\qquad& \Box^{(4)}\hat{v}_{\mu}=0, \nonumber\\
\text{tensor:} &\qquad& \Box^{(4)}\hat{D}_{\mu\nu}+\hat{D}_{\mu\nu}''-\frac{(a^{\frac{3}{2}})''}{a^{\frac{3}{2}}}\hat{D}_{\mu\nu}=0, \nonumber\\
\text{scalar:} &\qquad& \Box^{(4)}\hat{\mathcal{G}}+\hat{\mathcal{G}}''-\frac{\theta''}{\theta}\hat{\mathcal{G}}=0.
\end{eqnarray}

Since the vector perturbations only have zero modes, solutions are stable against vector perturbations \cite{YZYX}. However, for tensor and scalar modes another decomposition should be introduced \cite{YZYX}, namely,
\begin{eqnarray}
\hat{D}_{\mu\nu}(x^{\lambda},z)&=&\epsilon_{\mu\nu}e^{ip_{\lambda}x^{\lambda}}\rho_{p}(z), \nonumber\\
\hat{\mathcal{G}}(x^{\lambda},z)&=&e^{iq_{\lambda}x^{\lambda}}\Phi_{q}(z),
\end{eqnarray}
where $\epsilon_{\mu\nu}$ is the $TT$ polarization tensor and $\rho_{p}(z)$ and $\Phi_{q}(z)$ satisfy the following equations:
\begin{equation}\label{t}
\mathcal{A}_{t}\mathcal{A}_{t}^{\dag}\rho_{p}=m_{p}^{2}\rho_{p},
\end{equation}
\begin{equation}
\mathcal{A}_{s}\mathcal{A}_{s}^{\dag}\Phi_{q}=M_{q}^{2}\Phi_{q},
\end{equation}
with $m_{p}^{2}=-p^{\mu}p_{\mu}$, $M_{q}^{2}=-q^{\mu}q_{\mu}$ and
\begin{eqnarray}
\mathcal{A}_{t}&=&\frac{d}{dz}+\frac{(a^{\frac{3}{2}})'}{a^{\frac{3}{2}}}, \nonumber\\
\mathcal{A}_{s}&=&\frac{d}{dz}+\frac{\theta''}{\theta}.
\end{eqnarray}

Equation (\ref{t}) is a Schr\"{o}dinger-like equation, and takes the form
\begin{equation}\label{sch1}
-\rho_{p}''+\mathcal{W}(z)\rho_{p}=m_{p}^{2}\rho_{p},
\end{equation}
where the effective potential $\mathcal{W}(z)$ is given
\begin{eqnarray}\label{W}
\mathcal{W}(z)&=&\frac{(a^{\frac{3}{2}})''}{a^{\frac{3}{2}}}=\frac{\left(e^{\frac{3}{2}A(z)}\right)''}{e^{\frac{3}{2}A(z)}},\nonumber\\
&=&\frac{3}{2}A''+\frac{9}{4}A'^{2}.
\end{eqnarray}
The effective four-dimensional gravity is determined by the spectrum of the tensor $KK$ modes $\rho_{p}$ \cite{YZYX}. For instance, for zero mode $\rho_{0}$ with $m_{0}=0$ and a normalizable $\rho_{0}$ leads to the four-dimensional Newton's law \cite{Randall:1999vf,YZYX,41}.

\begin{figure}[h!]
\epsfxsize=11.5cm\centerline{\epsfbox{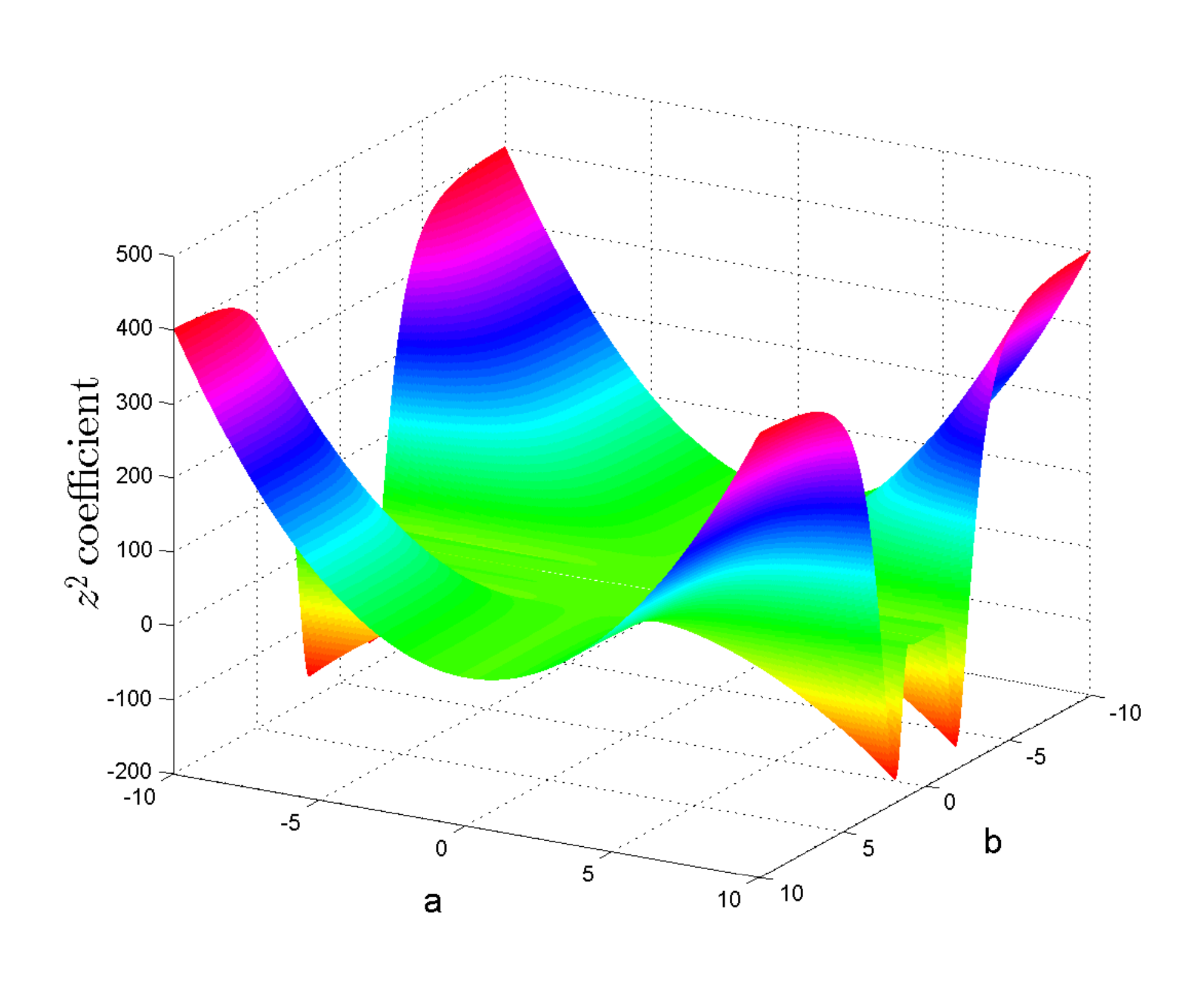}} \caption{The
coefficient of the $z^2$ term in the potential of the linearized
Schr\"{o}dinger equation as a function of the free parameters $a$
and $b$ for the SG system. Negative values correspond to a first order
instability.\label{OriginalSGz2}}
\end{figure}
\begin{figure}[h!]
\epsfxsize=11.5cm\centerline{\epsfbox{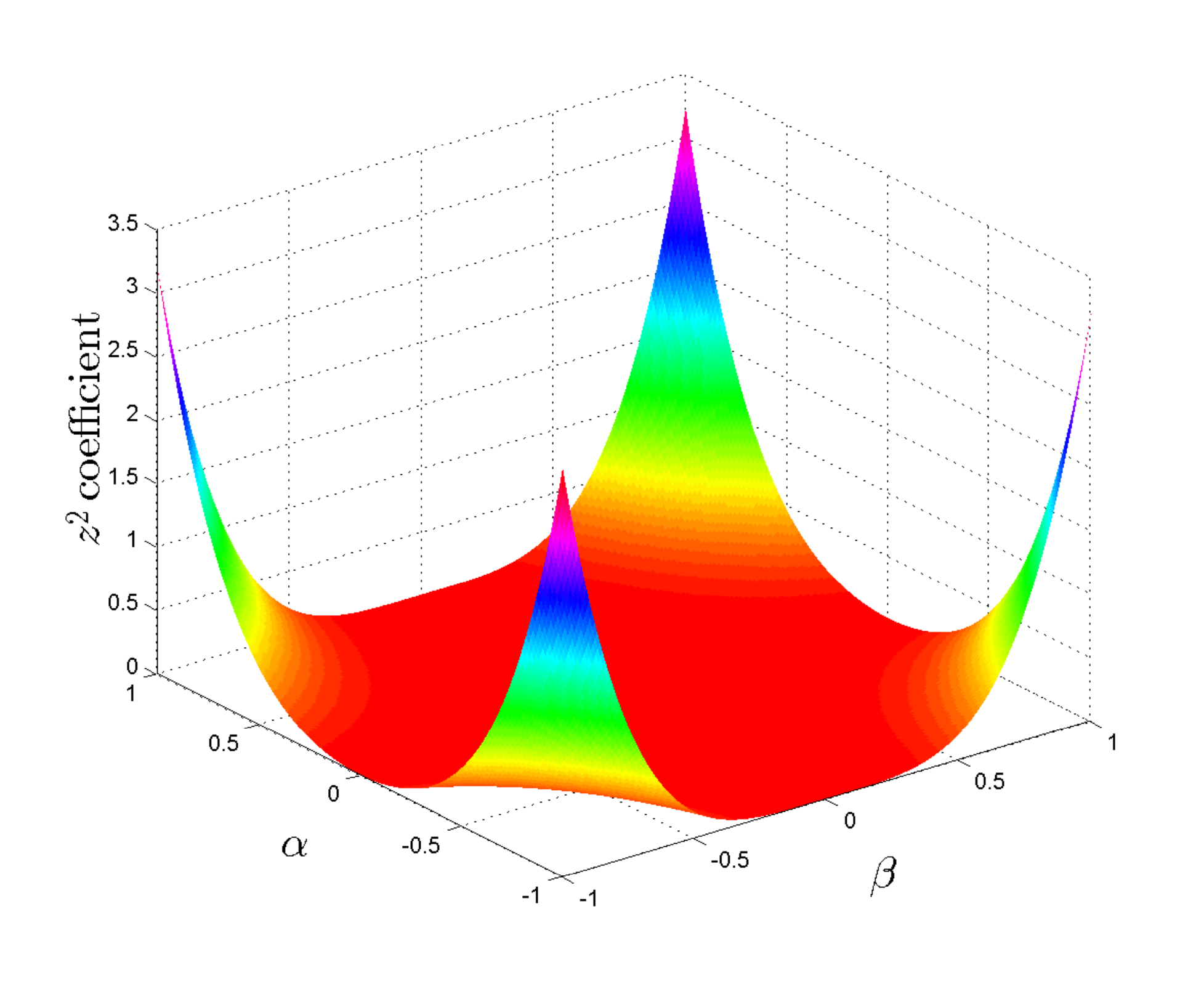}} \caption{The
coefficient of the $z^2$ term in the potential of the linearized
Schr\"{o}dinger equation as a function of the free parameters
$\alpha$ and $\beta$ for the $\phi^4$ system. The coefficient is
everywhere positive, signaling linear stability. \label{phi4z2}}
\end{figure}
\begin{figure}[h!]
\epsfxsize=11.5cm\centerline{\epsfbox{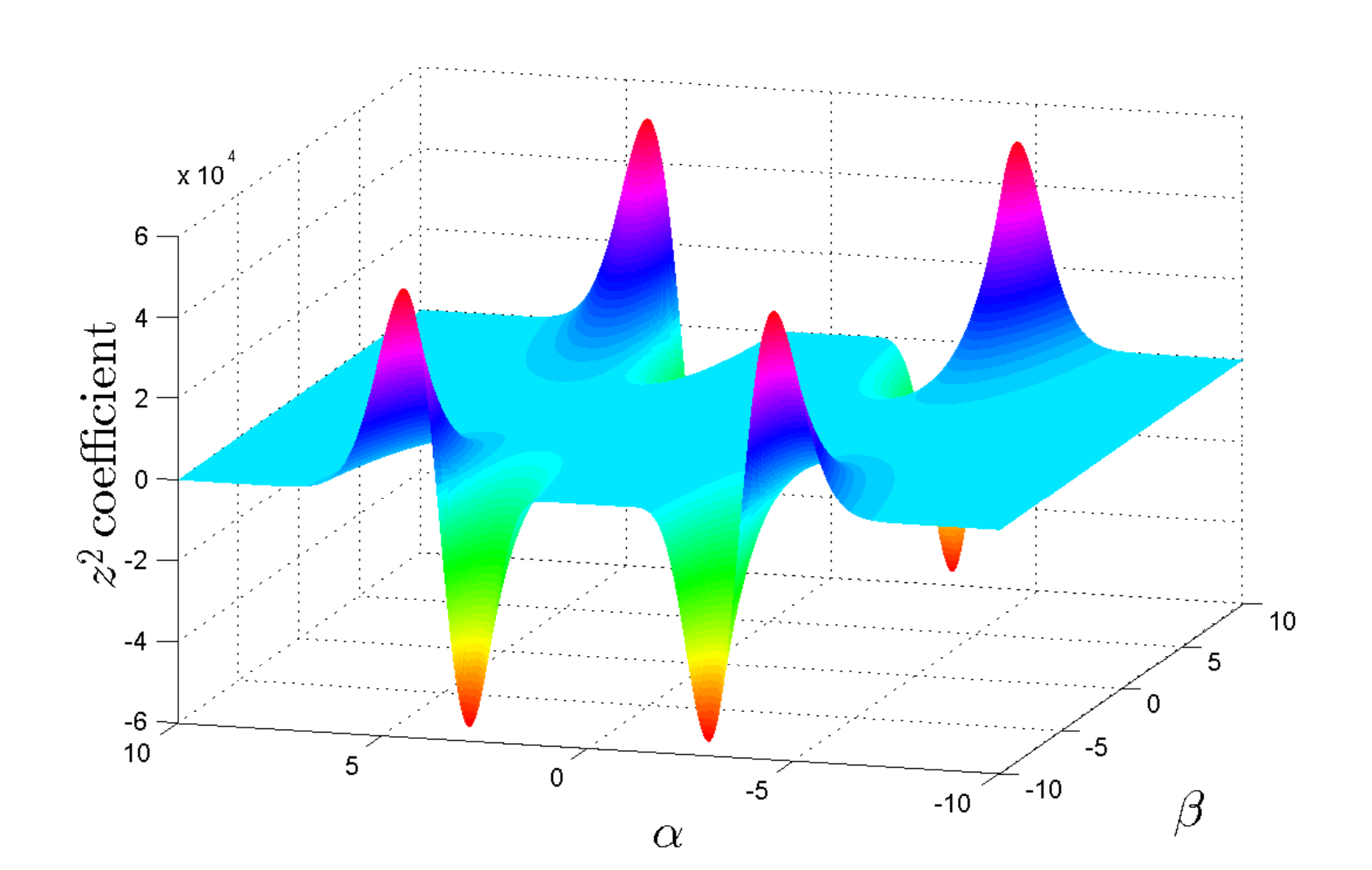}} \caption{The
coefficient of the $z^2$ term in the potential of the linearized
Schr\"{o}dinger equation as a function of the free parameters
$\alpha$ and $\beta$ for the $\phi^6$ system. Negative values
correspond to a first order instability. There are also vast patches
in the parameter space which have almost neutral stability.
\label{phi6z2}}
\end{figure}

On the other hand, in order to study the stability of the branes, some authors choose an ``axial gauge'' where the metric is perturbed as \cite{de,Sa,Af,CM,DA}:
\begin{equation}
ds^{2}=e^{ 2A(w)}(g_{\mu\nu} + \varepsilon h_{\mu\nu} )dx^{\mu} dx^{\nu} - dw^{
2}.
\end{equation}
Here $g_{\mu\nu}$ represents the background metric,
$h_{\mu\nu}$ denotes the metric perturbations, and $\varepsilon$ is a
small parameter \cite{Af}. They also consider the transformation $\varphi\longrightarrow\varphi+\varepsilon \tilde{\varphi}$, where $\tilde{\varphi}=\varphi(x,w)$ \cite{DA}. Moreover, in order to render the metric conformally flat, one can choose $dz=e^{-A(w)} dw$. In this case, the corresponding Schr\"{o}dinger equation takes the form \cite{de,AB,Sa,Af,DA,CM,BFG}:
\begin{equation}\label{sch}
-\frac{d^{2}\psi(z)}{dz^{2}}+U(z)\psi(z)=k^{2}\psi(z),
\end{equation}
where the potential is given by:
\begin{equation}
U(z)=-\frac{9}{4}\Lambda+\frac{9}{4}A'^{2}+\frac{3}{2}A''.
\end{equation}
As one can check that, this potential and the $\psi$ function are the same as $\mathcal{W}(z)$ and $\rho$ in Eq. (\ref{sch1}). Note that $\Lambda$ is a cosmological constant on the brane, which could be positive, negative or zero corresponding to the $4D$ spacetime being de Sitter ($dS_{4}$),
anti-de Sitter ($AdS_{4}$) or Minkowski ($M_{4}$)
\cite{Sa,Af}. Besides, it is notable that the Hamiltonian corresponding to Eq. (\ref{sch}) can be written in the form\cite{AB,Sa,BFG}:
\begin{equation}
H=\left(\frac{d}{dz}+\frac{3}{2}A'(z)\right)\left(-\frac{d}{dz}+\frac{3}{2}A'(z)\right),
\end{equation}
which is obviously Hermitian and therefore leads to real $k$ ($k^{2}\geq0$). Accordingly, there are no unstable tachyonic excitations in the system
\cite{DA,BFG}.

The solution for the zero modes ($k = 0$) is \cite{DA,YZYX,BFG}:
\begin{equation}
\psi(z)=Ne^{\frac{3A(z)}{2}}
\end{equation}
where $N$ is a normalization factor\cite{DA,YZYX,BFG} and satisfies:
\begin{eqnarray}
1&=&\int_{-\infty}^{+\infty}dz|\psi_{0}(z)|^{2}=N^{2}\int_{-\infty}^{+\infty}dze^{3A(z)}\nonumber\\
&=&\frac{N^{2}}{l}\int_{-\infty}^{+\infty}dye^{2A(y)},
\end{eqnarray}
where $y=lw$ is a dimensionless variable.
The asymptotic behavior of the solutions at large y are checked and the result is that only the $\varphi^{4}$ system has a normalizable zero mode and thus stable.
On the other hand, as in quantum mechanical systems, we may check for the stability of the system via the existence of a real frequency, in bound states.
Since the potentials for the three systems considered in this paper, are too complicated to be solved analytically, we found the corresponding ground
state eigenvalues via expansions in terms of the fifth coordinate $w$.
One can deduce the stability up to $O(w^2)$ by looking at the sign of the
$w^2$ term. It is seen in Figures \ref{OriginalSGz2}-\ref{phi6z2} that this coefficient is everywhere positive (stability) for the $\varphi^4$ system, while there are regions of the parameter space where the coefficient is negative (instability) for the $SG$
and $\varphi^6$ systems. However, as noted in section \ref{22a}, this conclusion is not decisive, since higher order effects might have drastic effects. The difference between the results of these two approaches is probably caused by the inevitable approximations used in the analysis.

\section{Conclusion}\label{Concl}

In this work, we obtained exact thick brane models inspired by
well-known nonlinear systems, namely, the sine-Gordon ($SG$),
$\varphi^{4}$ and $\varphi^{6}$ models. The confining effect of the
scalar field in all these three models were confirmed by examining
the geodesic equation for a test particle moving normal to the
brane. In particular, it turns out that the modified potential for
the $SG$ system resembles that of the double sine-Gordon ($DSG$)
system, while those of $\varphi^{4}$ and $\varphi^{6}$ became
$\varphi^{6}$ and $\varphi^{8}$, respectively.
We have extended previous brane models \cite{blg,br,CM} based on
$SG$ and $\varphi^4$ solitons taking into account different
parametrizations. The similarity of the $\varphi^4$ model with the
generic Higgs model makes this choice particularly interesting,
especially as the resulting potential is an odd function of the
fifth coordinate and the $Z_2$ symmetry is respected. We have
studied the $\varphi^6$ model for the first time.  This model is
interesting by its own right, since unlike the $\varphi^4$ model, we
have two pairs of solitons and anti-solitons which live in
different topological sectors.

In the case of the $SG$ model, we have used a more general form of
the potential compared to the one used in \cite{blg} and \cite{CM}.
The resulting brane does not have $Z_2$ symmetry, in general, where
the center of the brane may be displaced from $w=0$ and the
potential will not be an odd function of $w$ in general. However, by
a suitable choice of the model parameters it is possible to make the
vacuua of the effective potential degenerate, in which case the
$Z_2$ symmetry is restored. In the case of the $\varphi^6$ model,
however, we could not restore this symmetry via re-parametrization.
Finally, using standard procedures, we examined the stability of the
thick branes, by determining the sign of the $w^2$ term in the
expansion of the potential for the resulting Schrodinger-like
equation. It turns out that the $\varphi^4$ brane is stable, while
there are unstable modes for certain ranges of the model parameters
in the SG and $\varphi^6$ branes.

We considered the limiting case in
which the brane tends to zero thickness and approaches a thin brane.
It should be noted that the  topological stability of the soliton
brane remains valid even in this limit (at least at the classical
level). An interesting question would be whether the thick brane
continuous metric develops a discontinuity and whether the Israel
junction conditions will be satisfied in this limit. Although one
would expect intuitively that this is the case, we have not
worked out the detailed calculations. This issue will be explored
in separate paper.

\begin{acknowledgments}
N.R. acknowledges the support of Shahid Beheshti
University Research Council. M.P. acknowledges the support of
Ferdowsi University of Mashhad via the proposal No. 32361. F.S.N.L.
acknowledges financial  support of the Funda\c{c}\~{a}o para a
Ci\^{e}ncia e Tecnologia through an Investigador FCT Research
contract, with reference IF/00859/2012, funded by FCT/MCTES
(Portugal), and the grant EXPL/FIS-AST/1608/2013.
\end{acknowledgments}



\begin{thebibliography}{99}


\bibitem{Maartens:2003tw}
  R.~Maartens,
  ``Brane world gravity,''
  Living Rev.\ Rel.\  {\bf 7}, 7 (2004)
  [gr-qc/0312059].

\bibitem{Ge2}
R.~Davies, D.~P.~George and R.~R.~Volkas,
  ``The Standard model on a domain-wall brane,''
  Phys.\ Rev.\ D {\bf 77}, 124038 (2008)
  [arXiv:0705.1584 [hep-ph]].

\bibitem{lin}
G.~W.~Gibbons, R.~Kallosh and A.~D.~Linde,
  ``Brane world sum rules,''
  JHEP {\bf 0101}, 022 (2001)
  [hep-th/0011225].

\bibitem{PCU}
P.~Binetruy, C.~Deffayet, U.~Ellwanger and D.~Langlois,
  ``Brane cosmological evolution in a bulk with cosmological constant,''
  Phys.\ Lett.\ B {\bf 477}, 285 (2000)
  [hep-th/9910219].

\bibitem{Ge3}
D.~P.~George and R.~R.~Volkas,
  ``Kink modes and effective four dimensional fermion and Higgs brane models,''
  Phys.\ Rev.\ D {\bf 75}, 105007 (2007)
  [hep-ph/0612270].

\bibitem{de}
O.~DeWolfe, D.~Z.~Freedman, S.~S.~Gubser and A.~Karch,
  ``Modeling the fifth-dimension with scalars and gravity,''
  Phys.\ Rev.\ D {\bf 62}, 046008 (2000)
  [hep-th/9909134].

\bibitem{CG}
J.~A.~Cabrer, G.~von Gersdorff and M.~Quiros,
  ``Soft-Wall Stabilization,''
  New J.\ Phys.\  {\bf 12}, 075012 (2010)
  [arXiv:0907.5361 [hep-ph]].

\bibitem{Ge1}
D.~P.~George, M.~Trodden and R.~R.~Volkas,
  ``Extra-dimensional cosmology with domain-wall branes,''
  JHEP {\bf 0902}, 035 (2009)
  [arXiv:0810.3746 [hep-ph]].

\bibitem{Ge5}
D.~P.~George,
  ``Stability of gravity-scalar systems for domain-wall models with a soft wall,''
  J.\ Phys.\ Conf.\ Ser.\  {\bf 259}, 012034 (2010)
  [arXiv:1010.1628 [hep-th]].

\bibitem{blg}
P.~D.~Mannheim,
  ``Brane-localized gravity,''
  Hackensack, USA: World Scientific (2005) 337 p

\bibitem{Randall:1999ee}
  L.~Randall and R.~Sundrum,
  ``A Large mass hierarchy from a small extra dimension,''
  Phys.\ Rev.\ Lett.\  {\bf 83}, 3370 (1999)
  [hep-ph/9905221].

\bibitem{Randall:1999vf}
  L.~Randall and R.~Sundrum,
  ``An Alternative to compactification,''
  Phys.\ Rev.\ Lett.\  {\bf 83}, 4690 (1999)
  [hep-th/9906064].

\bibitem{Ge4}
D.~P.~George,
  ``Survival of scalar zero modes in warped extra dimensions,''
  Phys.\ Rev.\ D {\bf 83}, 104025 (2011)
  [arXiv:1102.0564 [hep-th]].

\bibitem{br} 
Damien P. George, 
\textit{Domain-wall brane models
of an infinite extra dimension}, Ph.D. Thesis, The University of
Melbourne, Australia (2009).

\bibitem{JY}
J.~Yang, Y.~L.~Li, Y.~Zhong and Y.~Li,
  ``Thick Brane Split Caused by Spacetime Torsion,''
  Phys.\ Rev.\ D {\bf 85}, 084033 (2012)
  [arXiv:1202.0129 [hep-th]].

\bibitem{RM}
R.~Menezes,
  ``First Order Formalism for Thick Branes in Modified Teleparallel Gravity,''
  Phys.\ Rev.\ D {\bf 89}, 125007 (2014)
  [arXiv:1403.5587 [hep-th]].

\bibitem{DBL}
D.~Bazeia, L.~Losano, R.~Menezes, G.~J.~Olmo and D.~Rubiera-Garcia,
  ``Thick brane in $f(R)$ gravity with Palatini dynamics,''
  arXiv:1411.0897 [hep-th].

\bibitem{OON}
O.~O.~Novikov, A.~A.~Andrianov and V.~A.~Andrianov,
  ``Gravity effects on the spectrum of scalar states on a thick brane,''
  PoS QFTHEP {\bf 2013}, 073 (2013).

\bibitem{Andrianov:2012ae}
  A.~A.~Andrianov, V.~A.~Andrianov and O.~O.~Novikov,
  ``Localization of scalar fields on self-gravitating thick branes,''
  Phys.\ Part.\ Nucl.\  {\bf 44}, 190 (2013)
  [arXiv:1210.3698 [hep-th]].


\bibitem{FD}
F.~Dahia and A.~de Albuquerque Silva,
  ``Classical tests of General Relativity in thick branes,''
  Eur.\ Phys.\ J.\ C {\bf 75}, no. 2, 87 (2015)
  [arXiv:1410.5463 [gr-qc]].

\bibitem{ALB}
A.~Ahmed, L.~Dulny and B.~Grzadkowski,
  ``Generalized Randall-Sundrum model with a single thick brane,''
  Eur.\ Phys.\ J.\ C {\bf 74}, 2862 (2014)
  [arXiv:1312.3577 [hep-th]].


\bibitem{AA}
V. Andrianov and A. A. Andrianov,
Proceeding of Science, The XXth International Workshop High Energy
Physics and Quantum Field Theory September 24 - October 1, (2011).

\bibitem{Andrianov:2013vqa}
  A.~A.~Andrianov, V.~A.~Andrianov and O.~O.~Novikov,
  ``Gravity effects on thick brane formation from scalar field dynamics,''
  Eur.\ Phys.\ J.\ C {\bf 73}, 2675 (2013)
  [arXiv:1306.0723 [hep-th]].

\bibitem{Ahmed:2013lea}
  A.~Ahmed, B.~Grzadkowski and J.~Wudka,
  ``Thick-Brane Cosmology,''
  JHEP {\bf 1404}, 061 (2014)
  [arXiv:1312.3576 [hep-th]].

\bibitem{AB}
V.~I.~Afonso, D.~Bazeia, R.~Menezes and A.~Y.~Petrov,
  ``f(R)-Brane,''
  Phys.\ Lett.\ B {\bf 658}, 71 (2007)
  [arXiv:0710.3790 [hep-th]].

\bibitem{BM}
D.~Bazeia, R.~Menezes and R.~da Rocha,
  ``A Note on Asymmetric Thick Branes,''
  Adv.\ High Energy Phys.\  {\bf 2014}, 276729 (2014)
  [arXiv:1312.3864 [hep-th]].

\bibitem{AIS}
A.~C.~Davis, S.~C.~Davis, W.~B.~Perkins and I.~R.~Vernon,
  ``Brane world phenomenology and the Z(2) symmetry,''
  Phys.\ Lett.\ B {\bf 504}, 254 (2001)
  [hep-ph/0008132].

\bibitem{DM}
D.~Yamauchi and M.~Sasaki,
  ``Brane world in arbitrary dimensions without Z(2) symmetry,''
  Prog.\ Theor.\ Phys.\  {\bf 118}, 245 (2007)
  [arXiv:0705.2443 [gr-qc]].

\bibitem{CC}
C.~Germani and C.~F.~Sopuerta,
  ``String inspired brane world cosmology,''
  Phys.\ Rev.\ Lett.\  {\bf 88}, 231101 (2002)
  [hep-th/0202060].

\bibitem{JL}
J.~E.~Lidsey,
  ``Inflation and brane worlds,''
  Lect.\ Notes Phys.\  {\bf 646}, 357 (2004)
  [astro-ph/0305528].

\bibitem{81}
Y.~B.~Zeldovich, I.~Y.~Kobzarev and L.~B.~Okun,
  ``Cosmological Consequences of the Spontaneous Breakdown of Discrete Symmetry,''
  Zh.\ Eksp.\ Teor.\ Fiz.\  {\bf 67}, 3 (1974)
  [Sov.\ Phys.\ JETP {\bf 40}, 1 (1974)].

\bibitem{86}
A.~Vilenkin,
  ``Gravitational Field of Vacuum Domain Walls and Strings,''
  Phys.\ Rev.\ D {\bf 23}, 852 (1981).

\bibitem{87}
A.~Vilenkin,
  ``Gravitational Field of Vacuum Domain Walls,''
  Phys.\ Lett.\ B {\bf 133}, 177 (1983).

\bibitem{88}
A.~Vilenkin,
  ``Cosmic Strings and Domain Walls,''
  Phys.\ Rept.\  {\bf 121}, 263 (1985).

\bibitem{Brihaye:2008am}
  Y.~Brihaye and T.~Delsate,
  ``Remarks on bell-shaped lumps: Stability and fermionic modes,''
  Phys.\ Rev.\ D {\bf 78}, 025014 (2008)
  [arXiv:0803.1458 [hep-th]].

\bibitem{Ahmed:2012nh}
  A.~Ahmed and B.~Grzadkowski,
  ``Brane modeling in warped extra-dimension,''
  JHEP {\bf 1301}, 177 (2013)
  [arXiv:1210.6708 [hep-th]].

\bibitem{Sa}
J.~Sadeghi and A.~Mohammadi,
  ``Shape invariance for the bent brane with two scalar fields,''
  Eur.\ Phys.\ J.\ C {\bf 49}, 859 (2007).

\bibitem{Af}
V.~I.~Afonso, D.~Bazeia and L.~Losano,
  ``First-order formalism for bent brane,''
  Phys.\ Lett.\ B {\bf 634}, 526 (2006)
  [hep-th/0601069].

\bibitem{CM}
W.~T.~Cruz, R.~V.~Maluf, L.~J.~S.~Sousa and C.~A.~S.~Almeida,
  ``Gravity localization in sine-Gordon braneworlds,''
  arXiv:1412.8492 [hep-th].

\bibitem{DA}
D.~Bazeia and A.~R.~Gomes,
  ``Bloch brane,''
  JHEP {\bf 0405}, 012 (2004)
  [hep-th/0403141].

\bibitem{JS}
S.~Jalalzadeh and H.~R.~Sepangi,
  ``Classical and quantum dynamics of confined test particles in brane gravity,''
  Class.\ Quant.\ Grav.\  {\bf 22}, 2035 (2005)
  [gr-qc/0408004].


\bibitem{YZYX} Y.~Zhong and Y.~X.~Liu
``Pure geometric thick $f(R)$-branes: stability and localization of gravity''
 arXiv:1507.00630 [hep-th].


\bibitem{FDCR} F.~ Dahia and C.~ Romero
``Confinement and stability of the motion of test particles in thick branes,''
  Phys.\ Lett.\ {\bf B 651}, 232, (2007)
  [gr-qc/0702011].


\bibitem{Ri}
N.~Riazi and A.~R.~Gharaati,
  ``Dynamics of sine-Gordon solitons,''
  Int.\ J.\ Theor.\ Phys.\  {\bf 37}, 1081 (1998).

\bibitem{Pey}
M. Peyravi,  A. Montakhab, N. Riazi and A. Gharaati,
``Interaction properties of the periodic and step-like solutions of the double-Sine-Gordon equation,''
Eur. Phys. J. {\bf B 72}, 269, (2009).

\bibitem{KS}
K.~Schnulle,
  ``Charged balanced black rings in five dimensions,''
  J.\ Phys.\ Conf.\ Ser.\  {\bf 372}, 012071 (2012).


\bibitem{gui}
M. Guidry, Gauge Field Theories, Wiley (2007).

\bibitem{hosein}
S.~Hoseinmardy and N.~Riazi,
  ``Inelastic collision of kinks and antikinks in the phi6 system,''
  Int.\ J.\ Mod.\ Phys.\ A {\bf 25}, 3261 (2010).

\bibitem{51}
Y.~Zhong and Y.~X.~Liu,
  ``Linearization of thick K-branes,''
  Phys.\ Rev.\ D {\bf 88}, 024017 (2013)
  [arXiv:1212.1871 [hep-th]].


\bibitem{41}
D.~Bazeia, A.~S.~Lob\~{a}o and R.~Menezes,
  ``Thick brane models in generalized theories of gravity,''
  Phys.\ Lett.\ B {\bf 743}, 98 (2015)
  [arXiv:1502.04757 [hep-th]].

\bibitem{BFG}
D.~Bazeia, C.~Furtado and A.~R.~Gomes,
  ``Brane structure from scalar field in warped space-time,''
  JCAP {\bf 0402}, 002 (2004)
  [hep-th/0308034].



\end{thebibliography}
\end{document}